\documentclass[
  journal=pasa,
  manuscript=research-paper,
  year=202X,
  volume=YY,
]{cup-journal}

\usepackage{amsmath}
\usepackage{amssymb}
\usepackage{microtype}
\usepackage{booktabs}
\usepackage{hyperref}
\usepackage{siunitx}
\newcommand{\lookdir}{\ensuremath{\hat{p}}}
\newcommand{\skydir}{\ensuremath{\hat{q}}}
\newcommand{\pbdir}{\ensuremath{\hat{b}}}
\newcommand{\elempat}{\ensuremath{D_n(\lookdir,\nu)}}
\newcommand{\tabpat}{\ensuremath{D_{\rm ta}(\lookdir,\nu)}}
\newcommand{\tabfull}{\ensuremath{T(\skydir;\pbdir,\lookdir,\nu)}}

\title{MWA tied-array processing V: Super-resolved localisation via amplitude-only maximum likelihood direction finding}

\author{Bradley~W.~Meyers}
\affiliation{Australian SKA Regional Centre (AusSRC), Curtin University, Bentley, WA 6102, Australia}
\alsoaffiliation{International Centre for Radio Astronomy Research (ICRAR), Curtin University, Bentley, WA 6102, Australia}
\email[B.~W.~Meyers]{bradley.meyers@curtin.edu.au}

\author{Arash~Bahramian}
\affiliation{International Centre for Radio Astronomy Research (ICRAR), Curtin University, Bentley, WA 6102, Australia}

\received {dd Mmm YYYY}
\revised  {dd Mmm YYYY}
\accepted {dd Mmm YYYY}
\published{DAY MONTH 202X}


\begin{document}

\begin{abstract}
Interferometric localisation of transients and pulsars via tied-array beam processing is challenging and can be limited by the native spatial resolution achievable by the instrument, especially at low frequencies and for compact interferometers.
Knowledge of the telescope primary and tied-array beam patterns allows the exploitation of the beam structures and the relationship to measured quantities, such as signal-to-noise ratio, through radio direction finding techniques.
The additional information provides a ``super-resolved'' localisation (i.e., where the precision is much better than the native spatial resolution) of a source when there are multiple detections in adjacent tied-array beams. 
We demonstrate this approach using the Murchison Widefield Array (MWA) and its voltage capture and tied-array processing capabilities, with a specific focus on how it benefits the on-going Southern-sky MWA Rapid Two-metre pulsar survey as it starts producing more candidates requiring follow-up.
Examples of localisations with previously discovered MWA pulsars which were subsequently localised via imaging with higher spatial resolution interferometers are used to validate the process, along with localisations of a sample of known pulsars to demonstrate the robustness of the method and its uncertainty estimation.
\end{abstract}

\section{Introduction}\label{sec:intro}
The tied-array beam (TAB) processing capabilities for Murchison Widefield Array \citep[MWA;][]{mwa1,mwa2} has evolved substantially over the past decade.
\citet{mwabf1} introduced the concept of tied-array beamforming in the context of MWA voltages, and \citet{mwabf2} verified that methodology is feasible at large offsets from zenith and produces sensible polarisation properties.
\citet{mwabf3} developed and verified an ``inverse polyphase filterbank'' (synthesis filter) approach to recover the 1.28\,MHz coarse channels from the recorded 10\,kHz voltages, making coherent dedispersion tractable and allowing a wider range of millisecond pulsar science.
Most recently, \citet{mwabf4} introduced a multi-pixel beamformer designed to aid in pulsar survey science, and demonstrated its usefulness in mitigating ionospheric bulk shifts of pulsars in the field of view.
The development of these technical processing capabilities made it possible to consider conducting a full southern-sky pulsar survey with the MWA by leveraging the voltage capture system \citep[VCS;][]{vcs,mwax}, leading to the inception of the the Southern-sky MWA Rapid Two-metre (SMART) pulsar survey \citep{smart1,smart2}.

During data acquisition periods for the SMART survey the MWA configured in its ``compact'' layout.
In this configuration the maximum element baseline is $B_{\rm max} \approx 740$\,m although given the centrally condensed element layout the effective characteristic baseline is $\tilde{B}_{\rm c} \sim 300$\,m.
At the central frequency of SMART (154\,MHz), this corresponds to a TAB width of $\lambda/\tilde{B}_{\rm c} \sim 20\text{-}30^{\prime}$ (depending on how baselines are weighted, how the bandwidth is averaged, etc.).
This makes the offline voltage processing tasks feasible for an all-sky survey, enlarging the TAB size and reducing the total number required to tessellate the observed field of view.
A feature of the compact configuration is the introduction of ``Hex'' clusters, where tiles are arranged in a hexagonal layout (see Fig.~1 of \citealt{mwa2}), adding many redundant baselines and subsequently modifying the synthesised beam pattern to be both wider (short baseline dominated) and inherit a hexagonal spatial distribution on the sky.

Given the wider TABs, it becomes difficult to localise pulsar candidates from SMART to the level required for follow-up with higher frequency and/or higher spatial resolution facilities.
Fast imaging (e.g., ${\lesssim}$100\,ms integrations) is technically feasible due to the versatility of voltage capture observations, but can be computationally expensive (especially since the TABs have already been created).
Image based pulsar or transient detection is nominally sensitive to a different population of objects than those probed by classical beamforming approaches \citep[e.g.,][]{sbs+23,sbs+24}, and also relies on the ability to detect the transient source in individual snapshot images, which can be challenging for moderate to weak sources.
The imaging approach is further complicated when the interferometric elements are closely spaced where one suffers from a wide synthesised beam and confusion-limited images.
It is possible for the brightest sources to achieve comparable or better localisation via imaging and cross-matching to calibrated astrometric survey positions at similar frequencies \citep[e.g.,][]{gleam,tgss,gleamx1}, but there is no guarantee that pulsars, or other fast-transients, will appear in these catalogues.
For the MWA, where the correlator does not run concurrently with the VCS, the trade-off favours TAB localisation using the voltage data, whereas for systems where the correlator and beamforming engines can run simultaneously it is will often be better to attempt image-based localisation (assuming the transient is detectable on the relevant accumulation timescale).

Strategic efforts utilising higher spatial resolution telescopes are the most directly effective at refining the candidate positions, where both imaging and standard high time resolution data acquisition modes can help to finalise localisation to the level of ${\sim}10^{\prime\prime}$ or better. 
This allows for a rapid convergence of a follow-up position for transient sources, or even a working timing solution for discovered pulsars, as demonstrated by \citet{sbs+21}.
Producing a timing solution for such new pulsars would otherwise require a dedicated observing campaign over the course of ${\sim}1$\,year in order to break the covariance between the unknown pulsar position and the rotational (and possible binary) properties.
Achieving a robust position as early as possible in the timing process can help to reveal binary motion and to constrain an orbital ephemeris which further reduces the amount of time required to fully model the pulsar rotation.
However, the required input to these follow-up observations is a best-estimate localisation from the MWA data available \citep[see Section~4.2 of][]{smart1}.
While there is occasionally sky coverage of the candidate from MWA observations in its extended configurations (where the TAB width is ${\sim}2^\prime$ at 154\,MHz), this is not generally the case.
Typically, one only has the information from the tessellated TAB detections of candidates from the initial SMART search process, or a re-processed TAB grid surrounding the nominal candidate position.

Moment invariants of digitised images can be used to calculate image centroids to sub-pixel precision, where each of the pixels (or pixel regions) are intensity weighted, however, they are biased due to undersampling of the field \citep[e.g.,][]{tc86,tft+06,vpr09}.
This approach can be translated directly to using TABs (as a ``pixel'') where the intensity weighting is provided by the detection metric, e.g., the signal-to-noise ratio (S/N). 
In this case a simplistic and identical TAB shape is often assumed, and struggles to produce meaningful uncertainties.
These downfalls can be mitigated by using a realistic TAB shape and sensitivity estimate (which could change across the sampled sky) for each grid pixel position, along with the corresponding detection metric, and an appropriately tailored direction-finding algorithm to arrive at a more robust position and uncertainty estimate.
We can achieve ``super-resolved'' localisation by incorporating such additional information, wherein the localisation and uncertainty regions are far smaller than the nominal TAB width.

The fundamental localisation technique has been demonstrated previously in similar circumstances for the Low Frequency Array (LOFAR), Australian Square Kilometre Array Pathfinder (ASKAP) and MeerKAT interferometers \citep{osw15,bsm+17,bcb+23}, but is also deeply connected with radio direction finding and frequency estimation methodologies \citep[e.g.,][]{capon69,music86,lipsky04}, where instead of having multiple physical detectors focused on different bearings, we generate our detector signals via digital tied-array beamforming from voltage data.
In practice, distributed islands of likelihood are usually identified, where the interferometric element layout (and, therefore, TAB shape) strongly impacts the complexity of the likelihood map (e.g., see Section~3.1 and Fig.~1 of \citealt{chime8rpts}).
Nevertheless, the maximum-likelihood amplitude-based direction finding algorithms are robust (assuming individual element gains are well-known/calibrated) and relatively simple to approach, making them attractive and effective tools for current and next-generation radio telescopes.

In \S\ref{sec:methods} we adapt and develop the framework used to localise a target by leveraging the MWA tied-array beam pattern in the context of amplitude-only maximum likelihood direction finding, and in \S\ref{sec:verification} we verify the approach.
In \S\ref{sec:followup} we discuss the impact this approach can have on localisation for MWA candidates.
Finally, in \S\ref{sec:summary} we summarise and touch on the implications of this method in terms of follow-up efficiency for SMART and MWA VCS science more generally.

\section{Methodology}\label{sec:methods}
While we focus on direct applications to the MWA, the methodology can in principle be applied to any similar interferometric array where there is the capability to re-process voltage data, or where clusters of tied-array beams are formed in real time. 
An example of the latter is the TRAnsients and PUlsars with MeerKAT (TRAPUM) survey \citep{trapum}, and their localisation software package {\sc SeeKAT} \citep{bcb+23}.
The underlying localisation method can be summarised in the following steps:
\begin{itemize}
    \item Simulate (or otherwise obtain) a tied-array beam pattern as a function of sky position and frequency. Outlined in \S\ref{sec:skyresp}. 
    \item Consolidate and assign the target source detection significance metrics (flux density, S/N, etc.) to each respective tied-array beam pattern. Outlined in \S\ref{sec:snrratios}.
    \item Using the ratio of combinations of the tied-array beam pattern and its corresponding detection metrics, form a pseudo-likelihood and, therefore, probability map. Outlined in \S\ref{sec:metric}-\ref{sec:probmap}.
    \item Identify the peak of the the probability map as the most-likely position.
    \item Calculate an error estimate based on the 2D distribution of the likelihoods. Outlined in \S\ref{sec:locerr}.
\end{itemize}
The manner in which these steps are achieved can vary between telescopes, and in the case of the MWA some further tempering of the probability maps are required due to the complex beam patterns.
The ASKAP localisation method approaches the problem similarly, but uses a Bayesian estimator rather than a maximum-likelihood estimator when forming the probability maps.
Fundamentally, we apply the same core steps for the MWA, but specifically consider the more complex beam patterns and how this method helps to mitigate what would otherwise be an untenable localisation follow-up scheme.

\subsection{Sky response pattern}\label{sec:skyresp}
Here we will provide a summary of the relevant antenna theory principles needed to construct the information required for localisation inference (see e.g., \citealp{balanis2016} for a thorough exploration).
\citet{mtb+17} and \citet{smart3} have used a similar approach in the context of tied-array beamforming to facilitate flux density calibration of MWA pulsar data, and \cite{sol+15} in the context of the MWA primary beam patterns.

\subsubsection{Calculating the array factor}\label{sec:arrayfactor}
Consider an array with $N$ non-interacting receiving elements, where the $n$th element has a sky-response pattern expressed as \elempat, with $\nu$ as the observing frequency and $\lookdir=(\vartheta,\varphi)$ as the desired pointing direction in horizontal coordinates, i.e., azimuth angle $\varphi$ and zenith angle $\vartheta$.
We then consider an impinging planar wave from the look-direction, which we express as 
\begin{equation}
    \psi_n(\lookdir,\nu) = \exp\left[i\frac{2\pi \nu}{c} (\hat{\mathbf{k}}_{\lookdir}\cdot\mathbf{r}_n)\right].
    \label{eq:planewave}
\end{equation}
Here, $\mathbf{r}_n = [x_n, y_n, z_n]$ represents the $n$th element position vector\footnote{The exact tile-set in operation need not be the same throughout an observing semester, therefore, the TAB shape can change with observing epoch.} and $\hat{\mathbf{k}}_{\lookdir} = [\sin\vartheta\cos\varphi, \sin\vartheta\sin\varphi, \cos\vartheta]$ is the unit wavevector corresponding to \lookdir.
We may then define the tied-array voltage pattern, \tabpat, as the weighted mean of the elements in response to that plane wave, 
\begin{equation}
    \tabpat = \frac{1}{N} \sum_{n=1}^{N} w_n \elempat\psi_n(\lookdir,\nu),
    \label{eq:tabpat}
\end{equation}
where $w_n$ is a weighting function that we are free to define, but which will ultimately act to ``steer'' the tied-array voltage pattern on the sky. 
If one assumes that $D_n$ is identical for each element in eq.~\eqref{eq:tabpat}, then we can write the element response (i.e., the telescope primary beam) as $D_{\rm el}$ for each antenna, and it becomes a multiplicative factor that can be brought outside of the summation. 
We can then express the TAB pattern as $\tabpat =  A(\lookdir,\nu) D_{\rm el}$, defining the ``array factor'' to be
\begin{equation}
    A(\lookdir,\nu) = \frac{1}{N} \sum_{n=1}^{N} w_n \psi_n(\lookdir,\nu).
    \label{eq:af}
\end{equation}
This describes the response toward \lookdir, however, the element and TAB patterns are well defined at all points on the sky above the horizon, $\skydir = (\theta,\phi)$ for $0 \leq \theta \leq \pi/2$ and $0 \leq \phi < 2\pi $.
To compute the response at an arbitrary sky direction, \skydir, we set $w_n$ to be the complex conjugate of an impinging planar wave, eq.~\eqref{eq:planewave}, from the direction of \skydir, i.e.,
$w_n = \psi_n^\dagger(\skydir,\nu)$.
Therefore, for any sky position, eq.~\eqref{eq:af} becomes
\begin{equation}
    A(\skydir; \lookdir,\nu) = \frac{1}{N} \sum_{n=1}^{N} \psi_n^\dagger(\skydir,\nu) \psi_n(\lookdir,\nu).
    \label{eq:af_norm}
\end{equation}
This weighting scheme has the effect of ensuring that the squared-modulus, $C(\skydir; \lookdir,\nu) = |A(\skydir; \lookdir,\nu)|^2$, is constrained to the range $[0, 1]$ and is unity if $\lookdir \equiv \skydir$. 
It can also conveniently be treated as a proxy for the fractional sensitivity of a TAB pointing to some arbitrary sky location, \skydir, with respect to the look-direction, \lookdir. 

From the perspective of a beamforming operation, $C(\skydir; \lookdir,\nu)$ represents regions on the sky from which any signal will be included in the total intensity dynamic spectrum of the beam formed towards \lookdir.
The look-direction (and thus, the TAB pattern) is time-dependent for astrophysical sources in observations of any significant duration, i.e., $\lookdir \rightarrow \lookdir(t)$, but for notation simplicity we will keep the time dependence implicit until it becomes pertinent.

\begin{figure}[tb]
    \centering
    \includegraphics[width=0.95\linewidth]{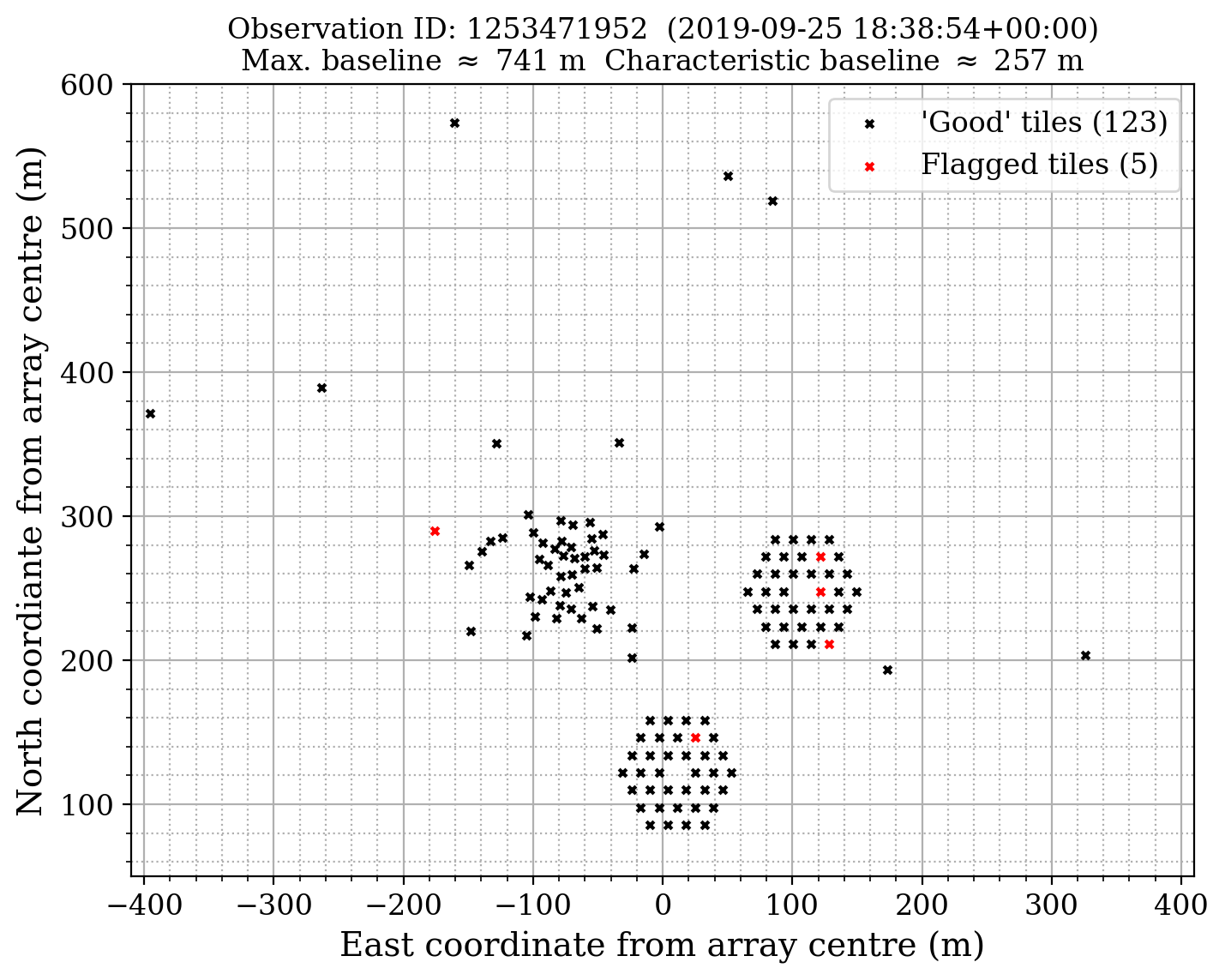}
    \includegraphics[width=0.95\linewidth]{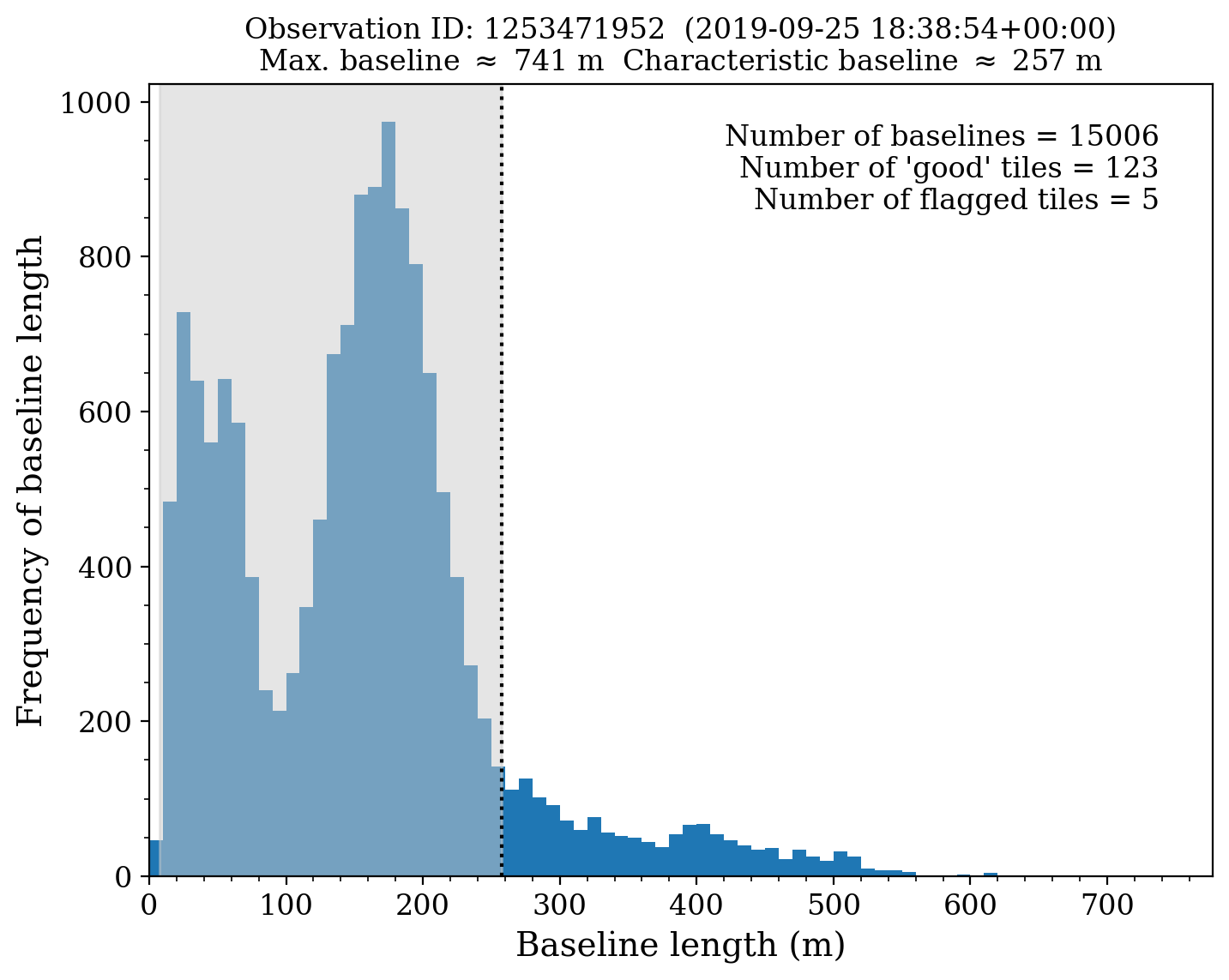}
    \caption{\textit{Top}: An example of a typical MWA compact configuration tile layout for Observation ID 1253471952, observed on 2019-09-25 18:38:54 UTC. 
    Here, each black cross is an operational element, or ``tile'', and red crosses indicate flagged (bad) tiles. 
    Usually, only 128 tile signals were recorded at any given time.
    \textit{Bottom}: The baseline length distribution, identifying the characteristic baseline length as ${\sim}260$\,m (black dotted line) as well as the 90\% highest density interval (grey shaded regions), i.e., the baseline length range that contain 90\% of all baselines. 
    }
    \label{fig:compact_layout}
\end{figure}

\begin{figure}[tb]
    \includegraphics[width=\linewidth]{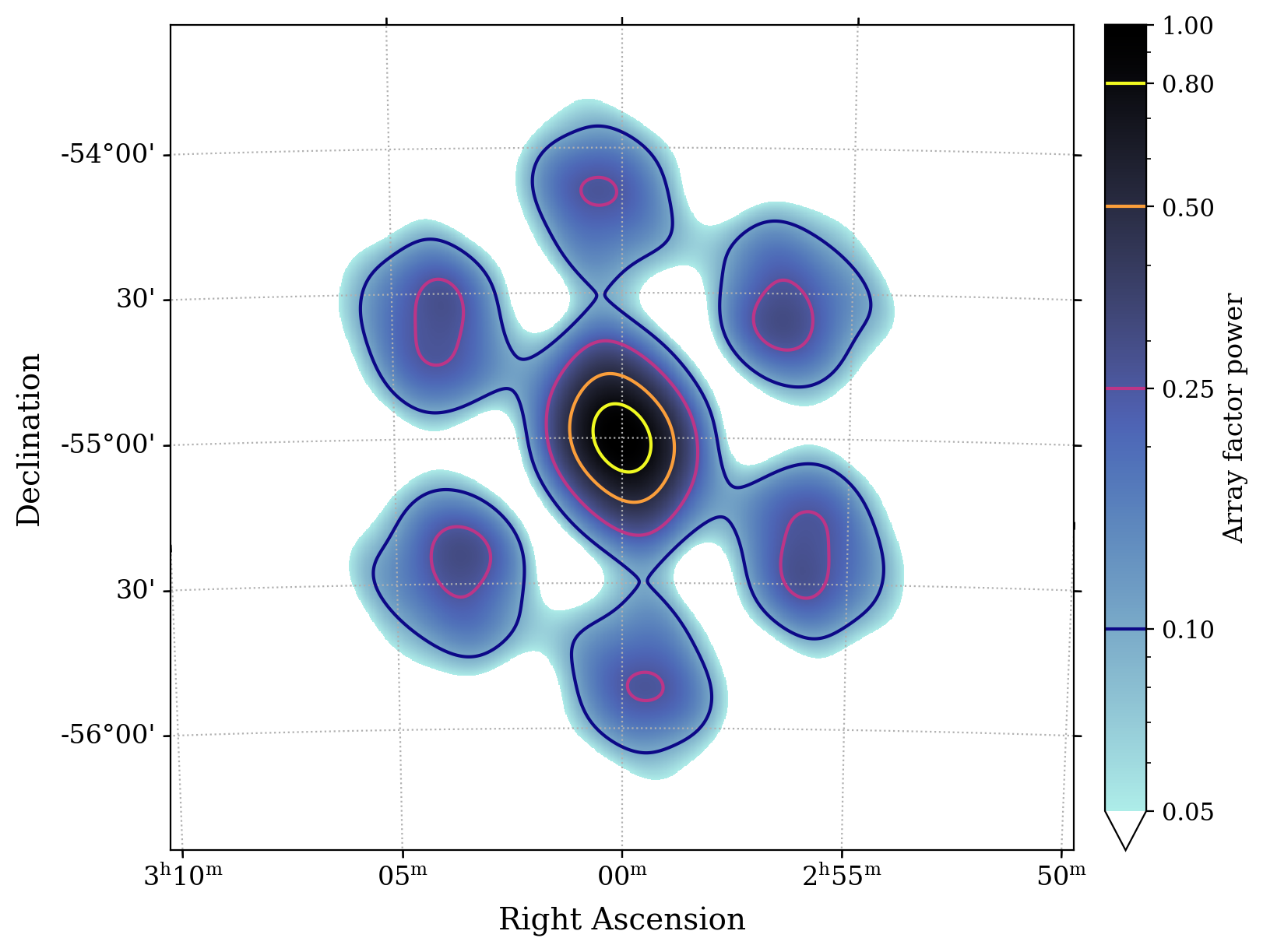}
    \includegraphics[width=\linewidth]{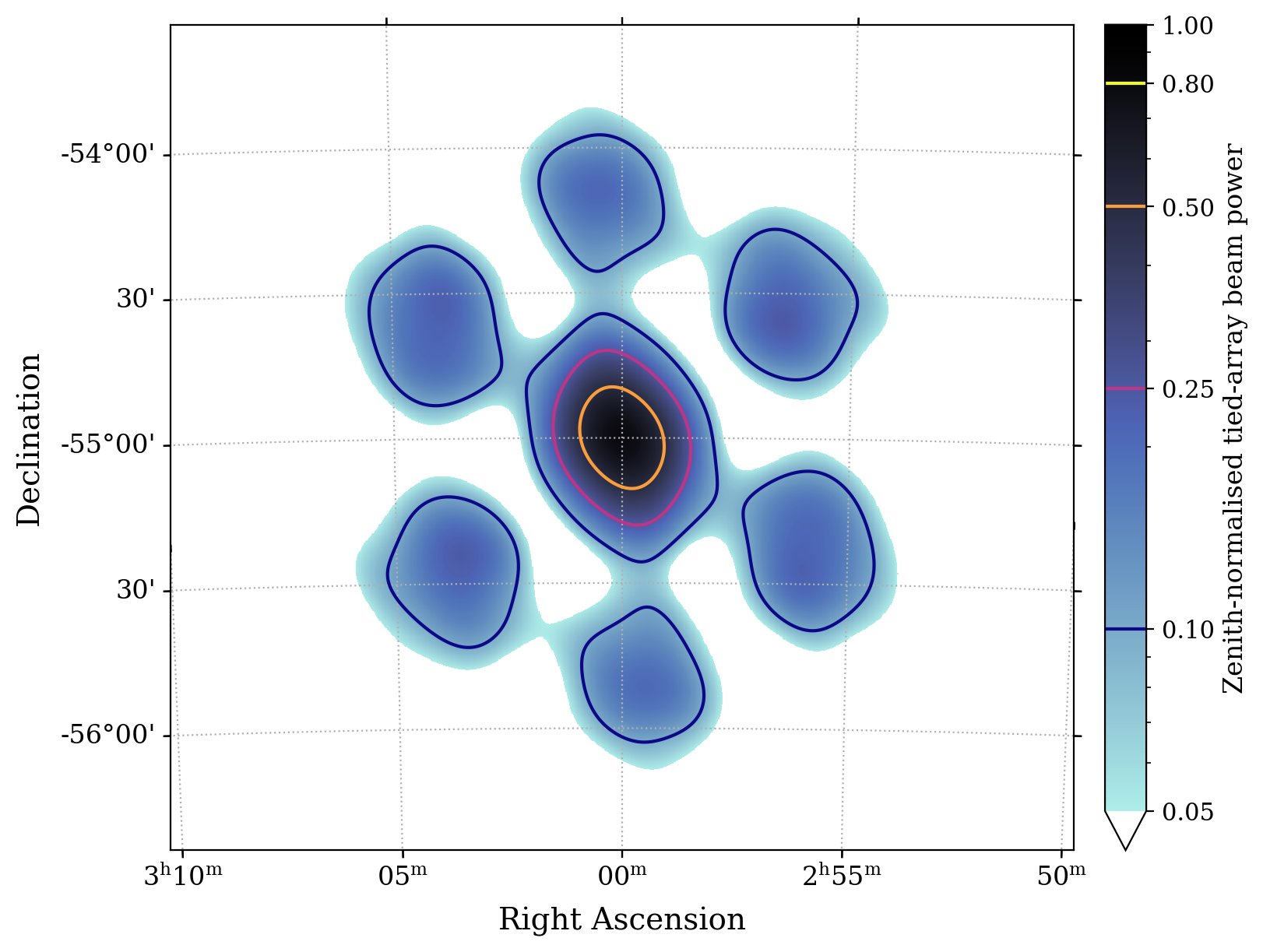}
    \caption{
    The tied-array beam pattern at 154\,MHz based on the element layout in Fig.~\ref{fig:compact_layout}.
    \textit{Top}: A zoom-in of the array factor power pattern, centred on the pointing target direction (approximately 30$^\circ$ off zenith). The effects of the compact array layout are visible in the first side-lobe patterns; a rotated and slightly asymmetrically hexagonal pattern.
    \textit{Bottom}: The same as the top panel, but it including the primary beam effects and normalised such that the primary beam power is unity at zenith. The target direction is just outside the 80\% zenith-normalised power point for this observation.}
    \label{fig:tab_pattern}
\end{figure}

An example of the MWA's compact configuration layout and baseline distribution with 128 tiles is given in Fig.~\ref{fig:compact_layout}, which produces the array factor power pattern as shown in the upper panel of Fig.~\ref{fig:tab_pattern}.

\subsubsection{Applying the primary beam apodisation}\label{sec:pbeffect}
To retrieve a realistic TAB response to the sky we must include the element pattern response, which can be achieved through multiplying $C(\skydir; \lookdir,\nu)$ by the squared-modulus of the element pattern, often referred to as the primary beam, $B(\skydir; \lookdir,\nu) = |D_{\rm el}(\skydir; \lookdir,\nu)|^2$. 

Initially, we have assumed that the element pattern and tied-array pattern are both pointed in the same direction.
This is not necessary, nor is it typical for beamformed observations with telescopes with static primary beam patterns.
The primary beam may be pointed to any arbitrary sky position, $\pbdir=(\theta_{\rm pb}, \phi_{\rm pb}$), and is usually static in horizontal (telescope) coordinates, whereas the TAB pattern is the desired ``pixel'' to evaluate for maximum sensitivity and usually moves across the sky with time (i.e., tracking a celestial source will result in a changing azimuth and zenith angle for all sources). 
Thus, the primary beam pattern transforms via $B(\skydir; \lookdir,\nu) \rightarrow B(\skydir; \pbdir,\nu)$ in the general case.

For the MWA, we use the \textsc{mwa\_hyperbeam} package\footnote{A Rust implementation with Python and C/C++ bindings. The pre-compiled binaries and source code can be found at \url{https://github.com/MWATelescope/mwa_hyperbeam}, or the Python package can be retrieved from PyPI.} to calculate the Full Embedded Element model of the primary beam response \citep{mwa_fee}.
Specifically, \textsc{mwa\_hyperbeam} produces a $2\times 2$ complex Jones matrix, $\mathbf{J}$, which represents the polarimetric primary beam response for each provided sky position at a given frequency.
These Jones matrices can be converted into a power response for each Stokes parameter via a transformation with the polarisation measurement equation \citep[e.g.,][]{hamaker00,vanStraten04}.
In our case, as we are specifically concerned with the total intensity (Stokes I) response, the measurement equation reduces to
\begin{equation}
    B = \frac{1}{2}{\rm tr}\left(\mathbf{J}\mathbf{I}_2\mathbf{J}^{\rm H}\right),
    \label{eq:pb_matmul}
\end{equation}
where ${\rm tr}$ is the trace operator, $\mathbf{M}^{\rm H}$ is the Hermitian transpose of a matrix $\mathbf{M}$, and $\mathbf{I}_2$ is the $2\times 2$ identity matrix (see eq.~10 in \citealt{mwabf1}).

\subsubsection{Creating the tied-array beam power pattern}\label{sec:tab}
Combining the above primary beam response with the array factor response provides a complete description of the theoretical tied-array power pattern, \tabfull, for any given celestial source position, observing frequency, and array configuration, across the entire sky,
\begin{equation}
    \tabfull \equiv C(\skydir; \lookdir,\nu) B(\skydir; \pbdir,\nu).
    \label{eq:tabp}
\end{equation}
In radio aperture synthesis imaging parlance, this is essentially the \textit{naturally weighted} synthesised beam pattern. 
An example of the TAB pattern based on the element layout in Fig.~\ref{fig:compact_layout} can be seen in the lower panel of Fig.~\ref{fig:tab_pattern}.

As an aside, the benefits of post-correlation beamforming techniques \citep[e.g.,][]{rcp18,price24}, compared to the standard brute-force operations represented here, include: the ability to incorporate additional weighting/tapering schemes which can be applied to reduce the TAB pattern effects (much like in standard synthesis imaging procedures), and that it notionally uses only the cross-correlation terms which can reduce red noise.
The downside of post-correlation beamforming is the limited time sampling, which is often driven by the practicalities of ${\sim}N^2$ visibility production in real-time.
This alternate beamforming method allows for flexible balancing of TAB field-of-view and overall survey sensitivity limits (a vital consideration for large pulsar survey projects, especially for the next-generation interferometric radio telescope facilities).

\subsection{Statistical localisation map}\label{sec:locmap}
We adopt the ``TABLo'' statistical localisation approach presented by \citet{bcb+23}. 
For completeness, we outline the fundamental steps and note any challenges specific to the MWA. 
Under the assumption of near-uniformity among TAB sensitivity (in a local grid), the starting premise is that the ratio of S/N measurements for each TAB pair is equal to the ratio of the TAB pair's sensitivity.
Notionally this holds only if the primary beam does not change drastically over the sampled sky region, or that the primary beam response is included in the model TAB sensitivity pattern.
There is also the consideration of the packing and overlap of adjacent TABs around the target candidates, but how to optimally decide this is beyond the scope of the work here (and, generally speaking, not obvious).
The S/N measurements, or their proxies, must be generated consistently across TABs and not use dynamic optimisation algorithms (i.e., methods that search parameter spaces to optimise S/N for data sets independently will introduce a bias into the collection that is hard to model).

\subsubsection{Per-beam S/N ratios and covariance}\label{sec:snrratios}
First, we construct a vector of ratios of each TAB S/N with the denominator set to the maximum measured S/N among any TAB considered ($S_{0}$),
\begin{equation}
    \vec{S}=\frac{S_i}{S_{0}},\quad\quad i=1,...,K,
    \label{eq:snr_ratio}
\end{equation}
where $S_i$ is the S/N for TAB $T_i$, and $i$ enumerates all TABs excluding that which contributes $S_0$ (i.e., there are $K+1$ total S/N measurements).

Creating ratios in this manner imparts a (small) covariance between entries in $\vec{S}$. 
To characterise this for each element in $\vec{S}$, we simulate a normal distribution where the mean is the entry value itself and the variance $\sigma^2=1$ (by definition of S/N). 
Conducting this simulation $L$ times for each of the $K$ elements in $\vec{S}$ results in a $K\times L$ matrix, which may be used to construct a covariance matrix, $\mathbf{C}$.
Defining $\textrm{cov}(S_i,S_j)=\mathbf{C}_{ij}$ to be the S/N covariance between TAB $i$ and $j$, then the $K\times K$ covariance matrix elements are
\begin{equation}
    \mathbf{C}_{ij} = \frac{1}{L-1}\sum_{m=1}^{L} (S_{im}-\bar{S}_{im}) (S_{jm}-\bar{S}_{jm}),\quad\quad i,j = {1, ..., K},
    \label{eq:tab_cov}
\end{equation}
where $S_{km}$ is the element drawn from the $m$th iteration distribution corresponding to the $k$th element in $\vec{S}$, and $\bar{S}_{km}\approx S_k$ is the mean value of that distribution.

\subsubsection{Statistical metric estimation}\label{sec:metric}
Using the result of eq.~\eqref{eq:tabp} and the elements of the vector in eq.~\eqref{eq:snr_ratio}, a residual vector at each sky position, $\vec{R}(\alpha,\delta)$ is computed such that
\begin{equation}
    R_i = S_i - T_i,\quad\quad i=1,...,K,
\end{equation}
where $T_i$ is the corresponding TAB which provided the measurement $S_i$.
which then allows us to employ a generalised least squares approach to compute a statistical metric that can be used to generate a likelihood map as a function of celestial coordinates.
This metric is the $\chi^2$ statistic, the weighted sum of the squared deviations at each sky position\footnote{This is a tensor operation because the residual vector is a tensor of shape $(K, N_\alpha, N_\delta)$, where $N_{\alpha}$ and $N_{\delta}$ are the number of pixels in the R.A. and Dec. axes, respectively, and the transpose operation is only along the first dimension, i.e., the dimension corresponding to the TAB index.},
\begin{equation}
    \chi^2(\alpha, \delta) = \vec{R}^{T}\mathbf{C}^{-1}\vec{R}
\end{equation}
which may then be used to determine the best-fit sky localisation and corresponding uncertainty.
This approach is essentially a 2-D application of amplitude-only maximum likelihood radio direction finding \citep[e.g., $P_{\rm ML}$ from][]{music86}.

\subsubsection{Metric regularisation}\label{sec:metricreg}
The resulting $\chi^2$ map may have many apparently significant points.
This is especially true in the MWA's case when using the compact tile configuration, but will be less problematic with a more pseudo-random tile layout or when the field of view is smaller/tied-array beam shape is more regular.
There are several reasons for these statistical ambiguities, including a sparse power distribution in the TAB patterns across the sky (which introduces large variations under the ratio operation), and that we completely ignore the absolute brightness/real flux density of the signal (because it is not always guaranteed to exist, whereas the S/N is commonly available).
These two realities mean that a typical $\chi^2(\alpha, \delta)$ map have several minima yielding a degenerate and multi-modal localisation map.
To mitigate this, we can re-weight the statistic map based on expectations of how far a typical source of interest can be from the TAB centre and still be significantly detected.
However, the choice of weighting function is not always obvious. 
Regularisation in this sense suppresses information based on assumed knowledge.
We recommend always inspecting the probability maps without such rescaling as it can reveal, for instance, side-lobe associations.

We elect, by default, to re-weight the statistical map with the TAB pattern with the greatest S/N detection, i.e., $\chi^2 \rightarrow T_0 \chi^2$. 
This approach favours weaker but nearby (to the $T_0$ pointing direction within the grid of TABs) likely localisations over the ones that are significantly stronger but further away.
Essentially, this means that a signal detected in the primary lobe of the TAB would need to be commensurately brighter if it were instead located in a less-sensitive part of the TAB for it to be detected with the same S/N.
On the downside, this drastically reduces the effective sky-area that is considered when computing the localisation probability by essentially down-weighting side-lobe contributions of other TABs that do not overlap with those of $T_0$. (See~\ref{app1}.)

One could instead use a less restrictive regularisation such as a multivariate Gaussian weighting function centred at the position of $T_0$, with a width as some multiple of the maximum distance between any TAB pair.
This still favours the $T_0$ pointing direction but allows information from scales near or beyond the first side-lobes of multiple TABs to be incorporated.

\subsubsection{Localisation probability maps}\label{sec:probmap}
Given the underlying $\chi^2$ basis, the log-likelihood at any celestial position, $(\alpha, \delta)$ is,
\begin{equation}\label{eq:loglike_cont}
    \ln \mathcal{L}(\alpha, \delta) \propto -\frac{1}{2}\chi^2(\alpha, \delta),
\end{equation}
and, hence, the relative probability density at a given celestial position becomes
\begin{equation}\label{eq:prob_cont}
    {\rm Pr}(\alpha, \delta) \propto \mathcal{L}(\alpha, \delta) \left/ \int_{\alpha,\delta} \mathcal{L}(\alpha, \delta).\right.
\end{equation}
For our purpose, where the $(\alpha, \delta)$ parameter space has been discretised on a grid, eqs.~\eqref{eq:loglike_cont} and \eqref{eq:prob_cont} become
\begin{equation}\label{eq:loglike_disc}
    \ln \mathcal{L}[\alpha, \delta] \approx -\frac{1}{2}\chi^2[\alpha, \delta],
\end{equation}
and
\begin{equation}\label{eq:prob_disc}
    {\rm Pr}[\alpha, \delta] \approx\mathcal{L}[\alpha, \delta] \left/ \sum_{\alpha,\delta} \mathcal{L}[\alpha, \delta].\right.
\end{equation}
The best localisation is then identified from eq.~\eqref{eq:prob_disc} via the ${\rm argmax}$ operation.

There is an important caveat to the above approach, namely that using $\chi^2$ as a basis implicitly assumes that the observations encoded through $T_i$ and $S_i$ are independent of each other and at least close to being normally distributed.
Given the way in which the TAB data are formed (by applying different complex phase rotation to the same underlying voltage streams), this assumption is not accurate.
While we do estimate a covariance matrix for $\vec{S}$, this covariance is mainly between the measured signal strengths and not their spatial covariance.
Furthermore, another implicit assumption of the framework is that any of the $2\pi$\,sr of sky not simulated as part of the TAB patterns contributes negligibly to the observed signals.

\subsubsection{Localisation uncertainty estimate}\label{sec:locerr}
The uncertainty in the localisation is computed using a similar method to that described in \citet{bcb+23}, based on the Mahalanobis distance \citep{mahalanobis, anderson}, $d_M$, under the assumption of a nearly-Gaussian probability distribution. 
By convention we use the 5$\sigma$ equivalent $d_M$ to derive a positional uncertainty.

In practice, there may be multiple ``islands'' of probability above the desired $d_M$ threshold, however the island of interest is that which contains the peak probability density value.
To determine this and the uncertainty region, we first create a mask that highlights only pixels with a value above the $d_M$ threshold.
The mask is then passed into a feature identification and labelling algorithm\footnote{Using the \texttt{ndimage.label} function from \textsc{SciPy}, with the default structural element matrix.}, and the island containing the peak localisation pixel is selected.
The Euclidean distance is computed from the peak pixel to each other pixel in the island, and the maximum value is taken as the conservative symmetric localisation uncertainty.
This can be extended to any arbitrary $n\sigma$ interval by simply re-evaluating the $d_M$ and the corresponding contour boundaries.

Due to the probabilistic nature of the technique (e.g., random sampling of Gaussian distributions to estimate the covariances) the final localisation and uncertainty can also change despite with identical inputs. 
In part, the magnitude of this random shift is driven by the sky sampling grid, as is the absolute probability density value, hence we only make use of the relative probabilities and do not place any strong meaning on the actual numerical quantities resulting from eq.~\eqref{eq:prob_disc}.
If the grid is too coarse, then the localisation method is insensitive to the small-scale TAB shape which is crucial to distinguishing the relative TAB powers and, therefore, the final localisation.
If the grid is too fine, the compute and memory cost increases without meaningful improvement.\footnote{The scaling increases with the number of pixels, $N_{\alpha}N_{\delta}$.}
Ultimately, the localisation peak and corresponding uncertainty shift by ${\sim}$1 pixel equivalent width.
Furthermore, while the assumption of well-calibrated interferometric element gains is reasonable it is not perfect, thus there will also be some systematic bias due to residual differences (though nominally the same within a given observation).

The ``pixel scale'' will necessarily depend on the observing setup, decreasing as frequency and/or baselines increase.
Generally, the pixel scale should correspond to no larger than $\sim 1/10$ of the TAB FWHM which ensures oversampling and a dynamic range suitable for identifying the most restrictive uncertainty intervals. 
In the following analysis, we choose a pixel size of $5^{\prime\prime}$, corresponding to ${\sim}300$ pixels across the nominal $24^\prime$ TAB FWHM for SMART.

\section{Verification with known pulsar positions}\label{sec:verification}
Here we demonstrate the localisation method using pulsars with well-constrained positions, starting with two specific examples alongside a broader sample study to explore the general accuracy of the localisation method.
Generally speaking, the predicted pulsar position relative to its interferometrically imaged location is offset by angles similar to those expected from random source position shifts induced by ionospheric turbulence and refractive effects (${\sim}0.1$-$1^\prime$) in this frequency range \citep[e.g.,][]{amo+15,cmt+17,mwabf4,wmj22}. 
To be conservative, one can add in quadrature ${\sim}1^\prime$ to the quoted systematic uncertainty.

\begin{figure*}[!hbt]
    \centering
    \includegraphics[width=0.495\linewidth]{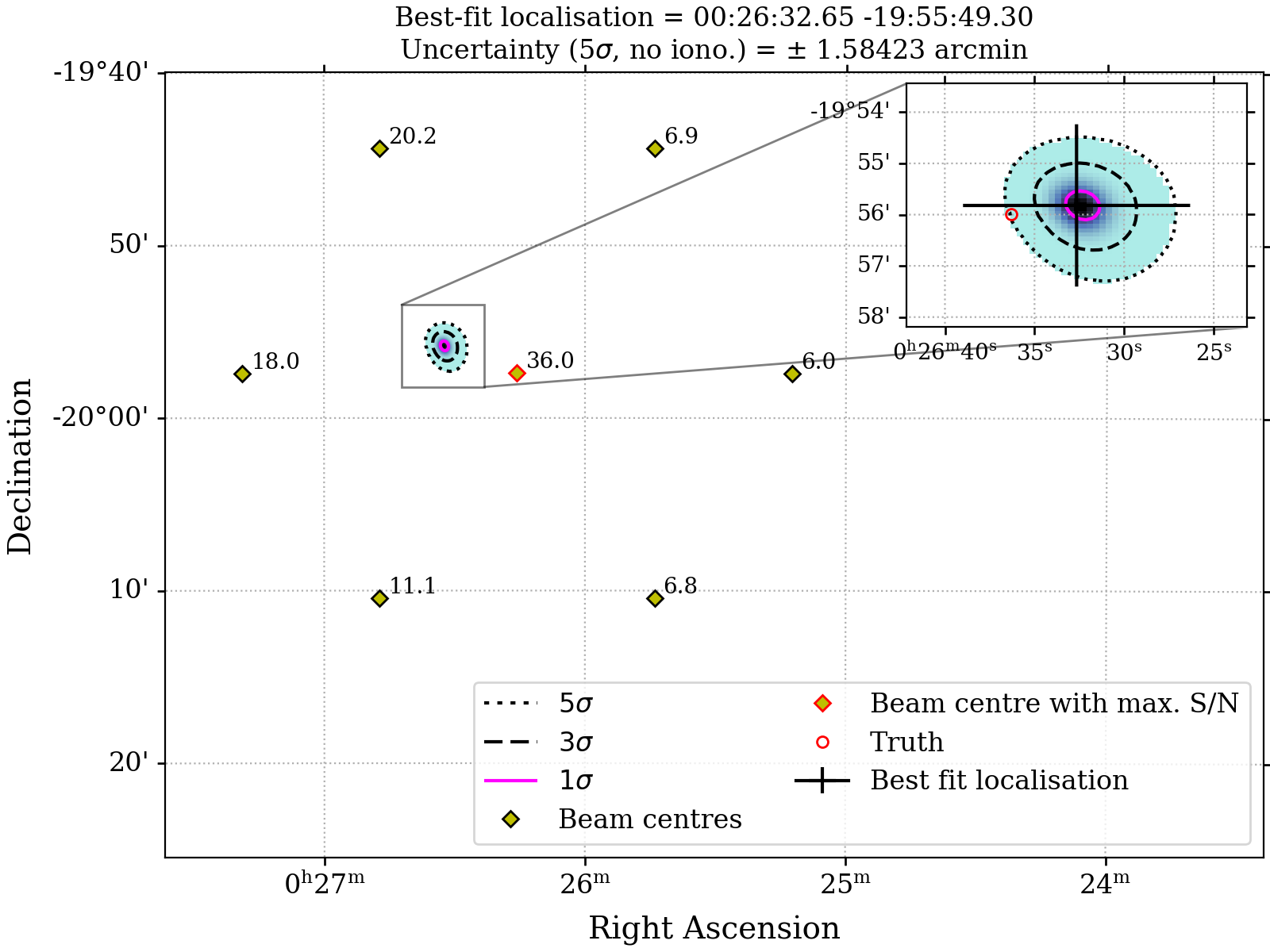}
    \includegraphics[width=0.495\linewidth]{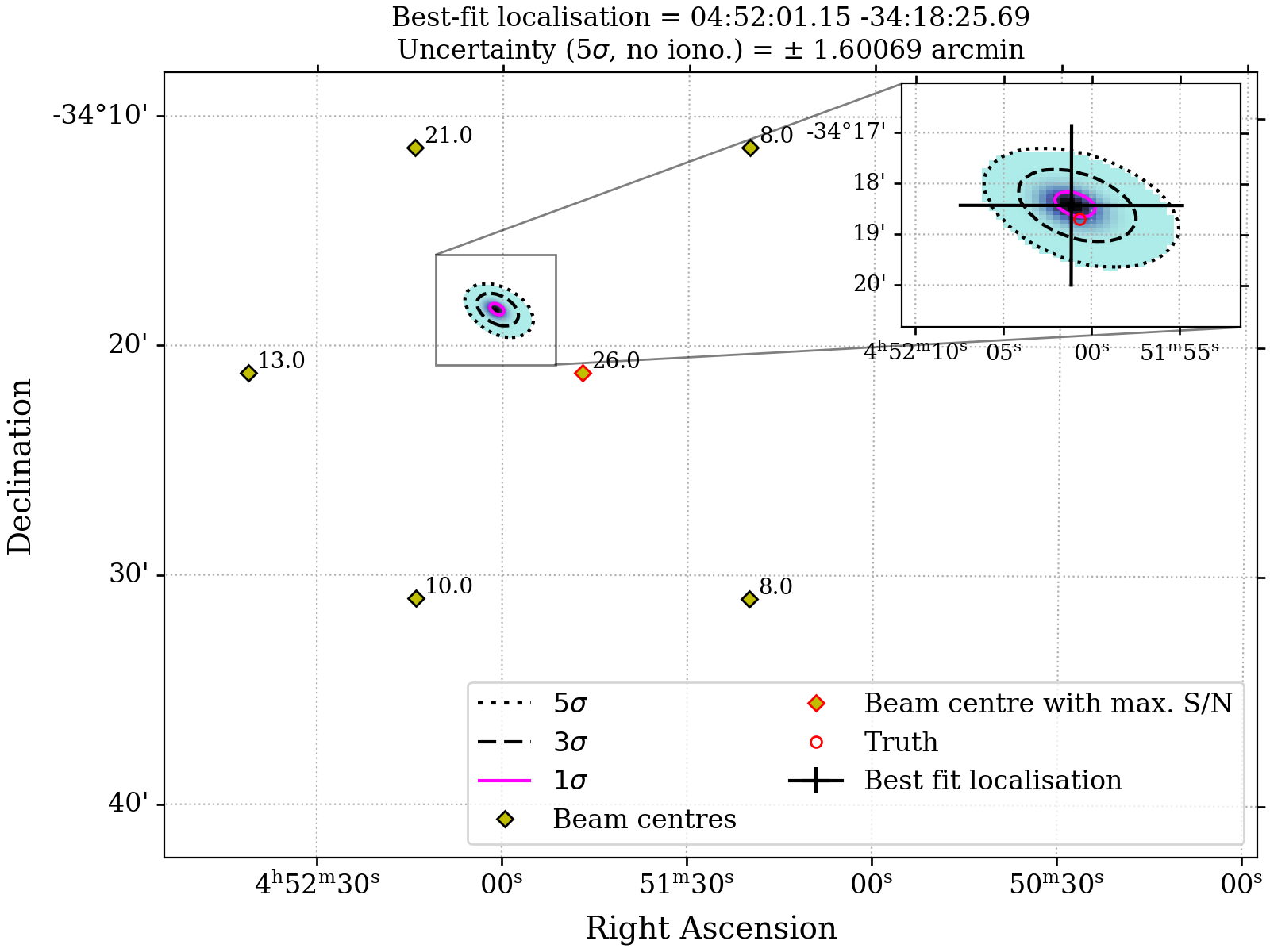}
    \caption{
    Localisation maps for two example SMART pulsar discoveries. 
    Each TAB in the grid their nominal pointing coordinates marked as a yellow diamond, with their corresponding measured S/N. 
    The TAB with the greatest S/N is also outlined in red, and the known pulsar position based on follow-up observations with higher-resolution telescopes (small red circle in the inset figure, labelled as ``Truth'') is reasonably well predicted. 
    Contours representing the 1, 3, and 5$\sigma$ uncertainties based on the probability density are overlaid, and the best-fit position with its corresponding symmetrical uncertainty is plotted in the inset.
    \textit{Left}: The localisation of PSR~J0026$-$1955 with the MWA using the original follow-up grid detections. 
    The 1, 3 and 5-$\sigma$ systematic uncertainties in this localisation instance are 0.3$^\prime$, 0.9$^\prime$, and 1.6$^\prime$ respectively, with a nominal offset from the known position of 0.87$^\prime$.
    \textit{Right}: The localisation of PSR~J0452$-$3418 with the MWA given the targeted grid detections (i.e., not the original SMART grid detections). 
    The 1, 3 and 5-$\sigma$ systematic uncertainties in this localisation instance are 0.3$^\prime$, 0.9$^\prime$, and 1.6$^\prime$ respectively, with a nominal offset from the known imaging position of 0.29$^\prime$.
    }
    \label{fig:psrs_loc}
\end{figure*}

\subsection{SMART PSR~J0026$-$1955}\label{sec:psr2}
The second discovered SMART pulsar (which was an independent re-discovery of the Green Bank North Celestial Cap (GBNCC) pulsar survey candidate J0025$-$19) went through a number of localisation iterations before being imaged \citep{mbs+22,smart2} with the upgraded Giant Metrewave Radio Telescope (uGMRT). 
This pulsar was in the minority where there where multiple archival MWA VCS data sets, including some with longer baseline configurations, which were used to constrain the pulsar position. 
Here, we demonstrate the localisation that would have been capable with just the initial SMART detections from 10-minutes of data in adjacent beams. 
We re-beamformed the data using the same grid setup as the original search data for 61 adjacent beams. 
Fig.~\ref{fig:psrs_loc} (left) demonstrates the MWA localisation from only the discovery observation.
The high angular resolution imaged pulsar position (in the J2000 epoch) is $\alpha_{\rm img} = 00^{\rm h}26^{\rm m}36^{\rm s}.3(2)$ and $\delta_{\rm img} = -19^{\circ} 55^{\prime}59^{\prime\prime}.3(3)$ \citep{smart2}.
Our localisation method predicts a position of $\alpha_{\rm loc} = 00^{\rm h} 26^{\rm m} 32^{\rm s}.65$ and $\delta_{\rm loc} = -19^{\circ} 55^{\prime} 49.30^{\prime\prime}$, with ($1\sigma$, $3\sigma$, $5\sigma$) uncertainties of ($0.3^\prime$, $0.9^\prime$, $1.6^\prime$) respectively, not including any ionospheric shift.
The two positions are offset by ${\sim}0.87^\prime$, thus match at a ${\sim}3\sigma$ level.

\subsection{SMART PSR~J0452$-$3418}\label{sec:psr5}
The fifth discovered SMART pulsar, PSR~J0452$-$3418 \citep{gbm+24}, was initially localised via a secondary (incomplete) gridded beamforming analysis (i.e., not the SMART gridded observations themselves) which was used to estimate the pulsar position. 
Targeted phased array beams from the uGMRT did not initially detect pulsations, thus the visibilities recorded in tandem were processed to produce synthesis images which ultimately localised the pulsar with a positional uncertainty of a few arcseconds. 
The original MWA localisation had a $3^\prime$ uncertainty, ignoring ionospheric effects, from a weighted centroid combination and was ultimately shown to be offset from the image-based position by ${\sim}2^\prime$.
The pulsar position (in the J2000 epoch), again from high angular resolution imaging, is $\alpha_{\rm img} = 04^{\rm h} 52^{\rm m}00^{\rm s}.7(3)$ and $\delta_{\rm img} = -34^{\circ} 18^{\prime} 42^{\prime\prime}(4)$ \citep{gbm+24,smart2}.
Fig.~\ref{fig:psrs_loc} (right) demonstrates the MWA localisation using the method, based on the same information initially used for the centroid-based refinement.
The predicted position is $\alpha_{\rm loc} = 04^{\rm h} 52^{\rm m} 01^{\rm s}.15$ and $\delta_{\rm loc} = -34^{\circ} 18^{\prime} 25^{\prime\prime}.69$, with ($1\sigma$, $3\sigma$, $5\sigma$) uncertainties of ($0.3^\prime$, $0.9^\prime$, $1.6^\prime$) respectively, not including any ionospheric shift. 
The positions are offset by ${\sim}0.29^\prime$, thus match at the ${\sim}1\sigma$ level.

\subsection{A sample of known pulsars}\label{sec:psr_samp_dist}
It is not straightforward to simulate and adequately capture the complexity of the tied-array beam and its impact on the S/N of fake targets. 
Instead, to approach the problem of general localisation accuracy, we produce several beamformed data sets of known pulsars in various places within the primary beam of the telescope and use these to empirically estimate the systematics involved.

We use a single observation (SMART G06; MWA Observation ID 1266329600) containing many detectable pulsars distributed across the primary beam, typically\footnote{There are a handful of pulsars that are barely detectable and/or extremely scattered, so we did not use those pulsars for this exercise owing to the complexity of defining a sensible S/N.} ranging from peak S/N of ${\sim}3$ to ${>}50$ when targeted directly. 
In total we created TAB grids around, and localised, 15 known pulsars.
For each pulsar position we applied an angular offset to it's known position in a random direction with a magnitude of 10-30$^\prime$ (i.e., between a half and a full SMART TAB width) and formed a hexagonal grid of overlapping TABs around the offset central position, similar to how the SMART sky tessellation is achieved.
Each TAB was generated from the full 80-minute drift scan data, processed with the corresponding and folded without S/N optimisation enabled.
The folding period and dispersion measure were gathered from the SMART data release provided by Bhat et al. (2026, submitted) and passed to the \texttt{prepfold} routine in the \textsc{PRESTO} software suite \citep{presto}.
For each TAB position the S/N was computed from the profiles by first identifying the ``on-pulse'' and ``off-pulse'' regions in the best detection, obtaining a noise estimate from the off-pulse region and comparing it to the maxima in the ``on-pulse'' region.

\begin{figure}
    \centering
    \includegraphics[width=\linewidth]{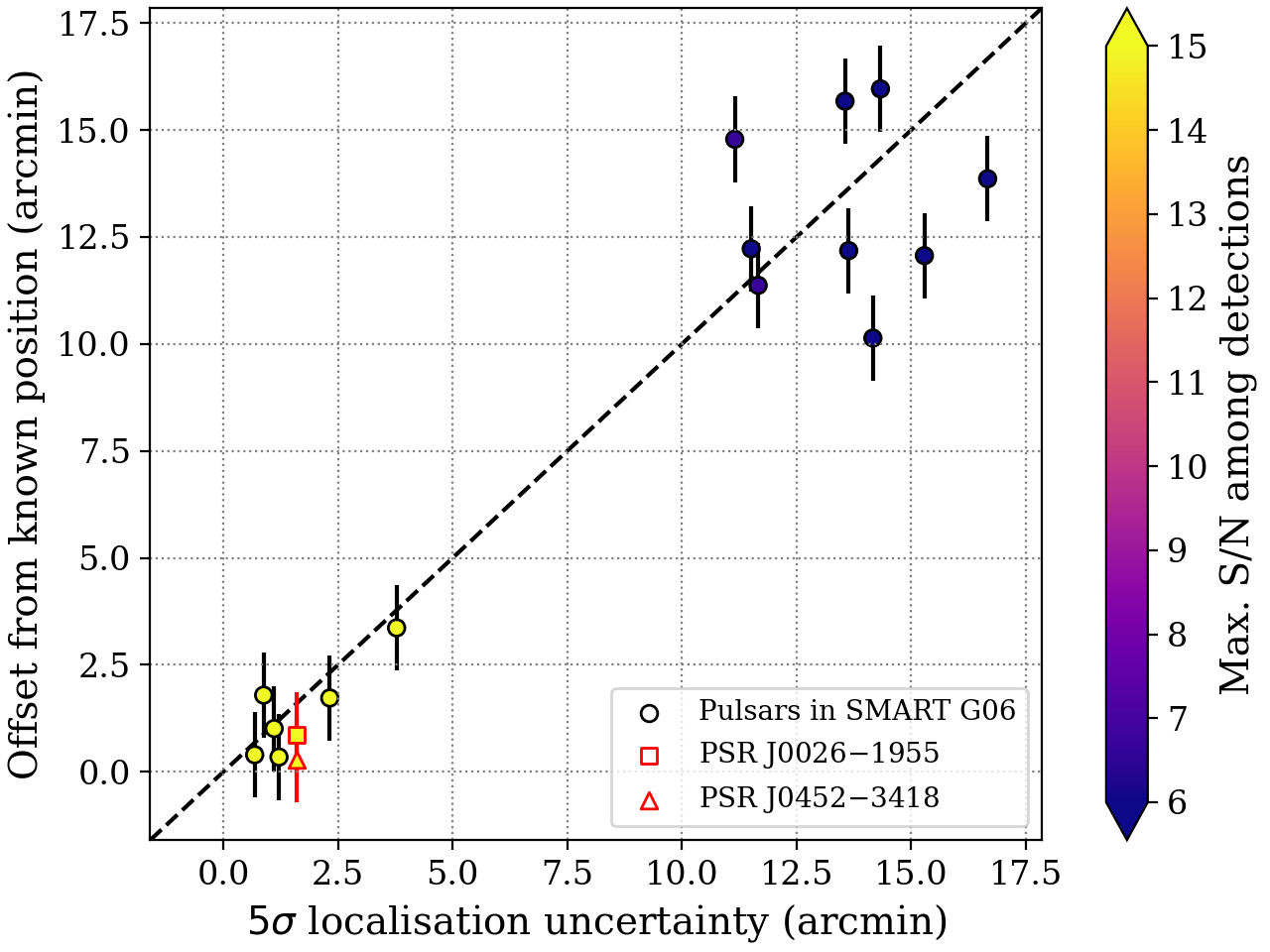}
    \caption{The localisation uncertainty vs. the offset of the known position from the best-fit localisation position for PSRs J0026$-$1955 (red outlined square), J0452$-$3418 (red outline triangle), and 15 detectable pulsars from the SMART G06 observation (circles). A one-to-one line is drawn for comparison. Generally, the localisation uncertainty matches any offset from the known pulsar position reasonably well, with increasing magnitude and spread as the S/N diminishes. Vertical error bars represent the maximum ionospheric refractive shift one might expect (${\sim}1^\prime$) in the known position vs. observed position.}
    \label{fig:offset_vs_error}
\end{figure}

In Figure~\ref{fig:offset_vs_error} we plot the offset from the known position of the pulsar and the best-fit localisation against the equivalent $5\sigma$ uncertainty.
Ideally, the data would lie on the equivalence line, or slightly below it, indicating that the uncertainty estimate encompasses the offset from the known position. This is the case for many of the examples, especially when including a conservative error contribution of ${\sim}1^\prime$ from ionospheric shifts. 
Many of the pulsar detections are marginal, and the range of S/N measurements across beams is small, which ultimately limits the information available and, therefore, the limits the localisation precision.
Broadly speaking the uncertainty region and the offset from known positions are relatively close, especially in the high S/N regime. 
As the detection S/N reduces, or if the diversity of S/N measurements over the TAB grid is small, the localisation effectiveness also reduces as the information provided loses constraining power.

\section{MWA pulsar and transient candidate follow-up}\label{sec:followup}
Instruments with a large native synthesised beam (and, therefore, tied-array beam) pattern compared to any plausible follow-up instrument incur a corresponding large overhead.
This is essentially due to the higher-resolution follow-up instrument needing to conduct a (smaller scale) search across the broader synthesised beam area. 
For SMART in particular this is challenging given the 20-30$^\prime$ TAB width and the available follow-up instruments with overlapping sky coverage.
In the case of single dish radio follow-up, Parkes/\textit{Murriyang} has a beam width $25.8\text{-}6.6^\prime$ with the Ultra Wide-bandwidth Low (UWL) receiver frequency coverage (0.7-4\,GHz), and the Green Bank Telescope has a beam width $25.1\text{-}3.6^\prime$ with the Ultra Wideband Receiver (UWBR) coverage (0.5-3.5\,GHz).
For interferometric follow-up, the uGMRT synthesised beam width is $3.5^\prime$ at 350\,MHz assuming only the central 1\,km antennas, while for MeerKAT it is $0.28^\prime$ at 600\,MHz with a maximum 8\,km baseline.

An illustrative example of this challenge is provided by \citet{gbm+24}, where multiple iterations of imaging and beamforming on subsequently detected targets was eventually required to localise PSR~J0452$-$3418.
Use of the methods described here would have saved much of the exploratory time (see \S\ref{sec:psr5}), which resulted in hours of uGMRT time being used to search for rather than observe the target.
Without localisation to a handful of arcminutes, targeted MeerKAT TAB gridding observations are expensive and/or infeasible. 
A recent SMART follow-up campaign with MeerKAT, which still required its own TAB gridding and localisation processing, led to a successful candidate localisation of PSR J0125$-$5854 (Tan et al. 2026, in prep.) based on the initial localisation provided by the method defined here.

\section{Summary}\label{sec:summary}
The method demonstrated here is an efficient way to obtain super-resolved localisation of events for telescopes: (1) when imaging and pulsar observing modes are not contemporaneous or easy to produce, (2) where multiple simultaneous TABs can be formed, and (3) when a reasonable model of the primary beam is available (since the TAB simply relies on interferometric element positions). 
The approach has been successfully adapted and demonstrated on at least two SKA precursors/demonstrators for both the Low and Mid telescopes, and may be useful for certain SKA science objectives and data processing tasks. 
Pathways worth exploring in future include: implementing subband localisation (where the frequency-dependence of the TABs may be exploitable to improve precision), and multi-epoch combinations (where the time-dependent refractive shifts from the ionosphere could be extracted).
Both trajectories would require special attention be paid to the ionosphere, and would help to quantify the effect on localisation given their time and frequency dependence. 
Additionally, with the Phase 3 expansion of the MWA (where all 256 tiles may be correlated/combined simultaneously, and a real-time beamforming mode is under development) the arcminute-level positions possible with SMART VCS observations and the described localisation technique provide an excellent starting point for MWA follow-up and position refinement of it's own candidates before moving to other instruments. 

For SMART, this approach will allow for rapid convergence on robust pulsar candidate positions, which will then accelerate follow-up observations with independent telescopes that have much smaller fields of view and/or are operating at higher frequencies.
It will also prove useful in localising single-pulse events, such as Rotating Radio Transients (RRATs), potential Fast Radio Burst candidates or Gamma-ray Burst prompt radio emission, which have both been quite elusive thus far at low frequencies.

\begin{acknowledgement}
This scientific work uses data obtained from Inyarrimanha Ilgari Bundara, the Murchison Radio-astronomy Observatory. 
We acknowledge the Wajarri Yamaji People as the Traditional Owners and native title holders of the Observatory site. 
Establishment of CSIRO's Murchison Radio-astronomy Observatory is an initiative of the Australian Government, with support from the Government of Western Australia and the Science and Industry Endowment Fund. 
Support for the operation of the MWA is provided by the Australian Government (NCRIS), under a contract to Curtin University administered by Astronomy Australia Limited. 

This work was supported by resources provided by the Pawsey Supercomputing Research Centre’s Setonix Supercomputer (\url{https://doi.org/10.48569/18sb-8s43}) and Garrawarla GPU Cluster (\url{https://doi.org/10.48569/gskb-tp15}), with funding from the Australian Government and the Government of Western Australia.

The authors thank Ramesh Bhat, Adrian Sutinjo and Sam McSweeney for useful discussions and suggestions throughout the development and writing process. 
We also thank the anonymous referee for their valuable feedback and suggestions.
\end{acknowledgement}

\paragraph{Software}
\textsc{Astropy} \citep{astropy:2013,astropy:2018,astropy:2022}, \textsc{NumPy} \citep{numpy}, \textsc{SciPy} \citep{scipy}, \textsc{Arviz} \citep{arviz_2019}, \textsc{Matplotlib} \citep{matplotlib}, \textsc{mwalib} (\href{https://github.com/MWATelescope/mwalib}{github.com/MWATelescope/mwalib}), \textsc{mwa\_hyperbeam} (\href{https://github.com/MWATelescope/mwa_hyperbeam}{github.com/MWATelescope/mwa\_hyperbeam}), \textsc{PRESTO} \citep{presto}.

\paragraph{Data Availability Statement}
The code developed here can be found on \href{https://github.com/CIRA-Pulsars-and-Transients-Group/mwa-vcs-localise}{GitHub} and is publicly available under the Academic Free License v3.
The data and example scripts used to generate the figures in this manuscript are also included.

\printendnotes

\bibliography{refs}

\appendix

\section{Effects of differing regularisation methods}\label{app1}
To demonstrate the effects of the different regularisation techniques, we take PSR~J0026$-$1955 as an example. 
In particular we focus on the original side-lobe detection of the pulsar \citep[see][]{mbs+22,smart2}. 
The pulsar was further localised after more candidates were collected from SMART data processing and it was realised that the first was a side-lobe detection.
The follow-up grid formed thereafter is what is replicated in \S\ref{sec:psr2}. 

\begin{figure}
    \centering
    \includegraphics[width=\linewidth]{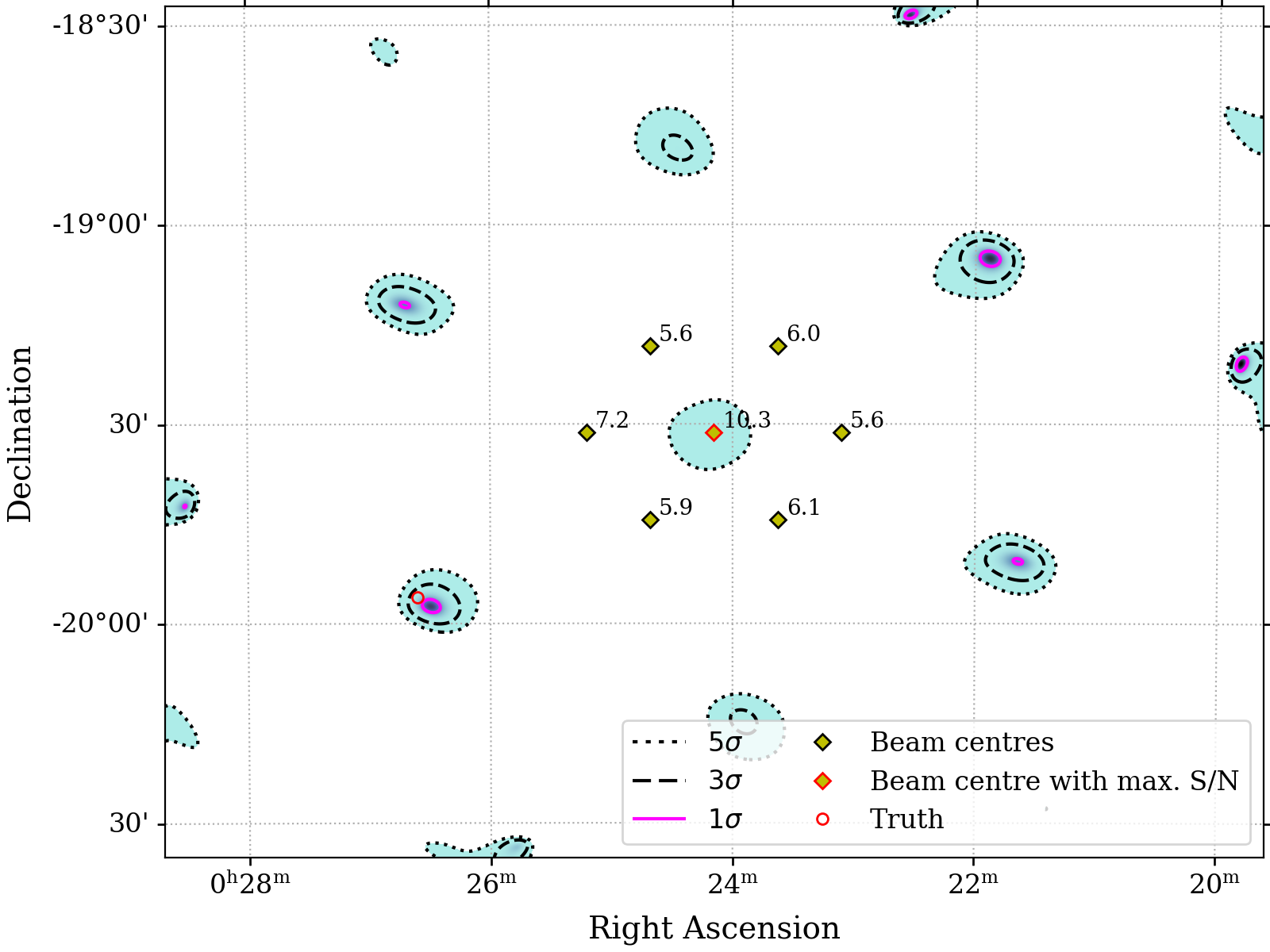}
    \includegraphics[width=\linewidth]{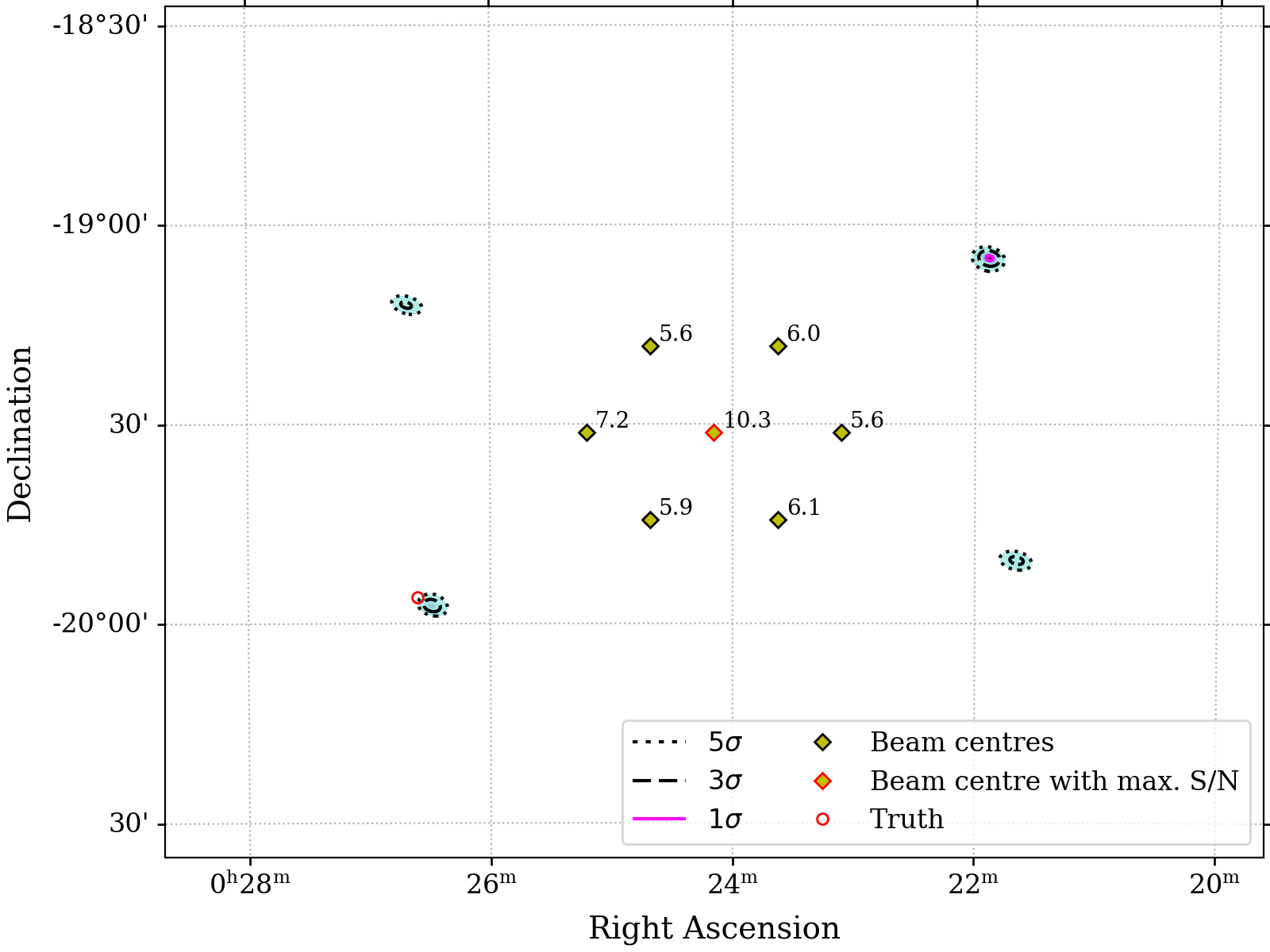}
    \includegraphics[width=\linewidth]{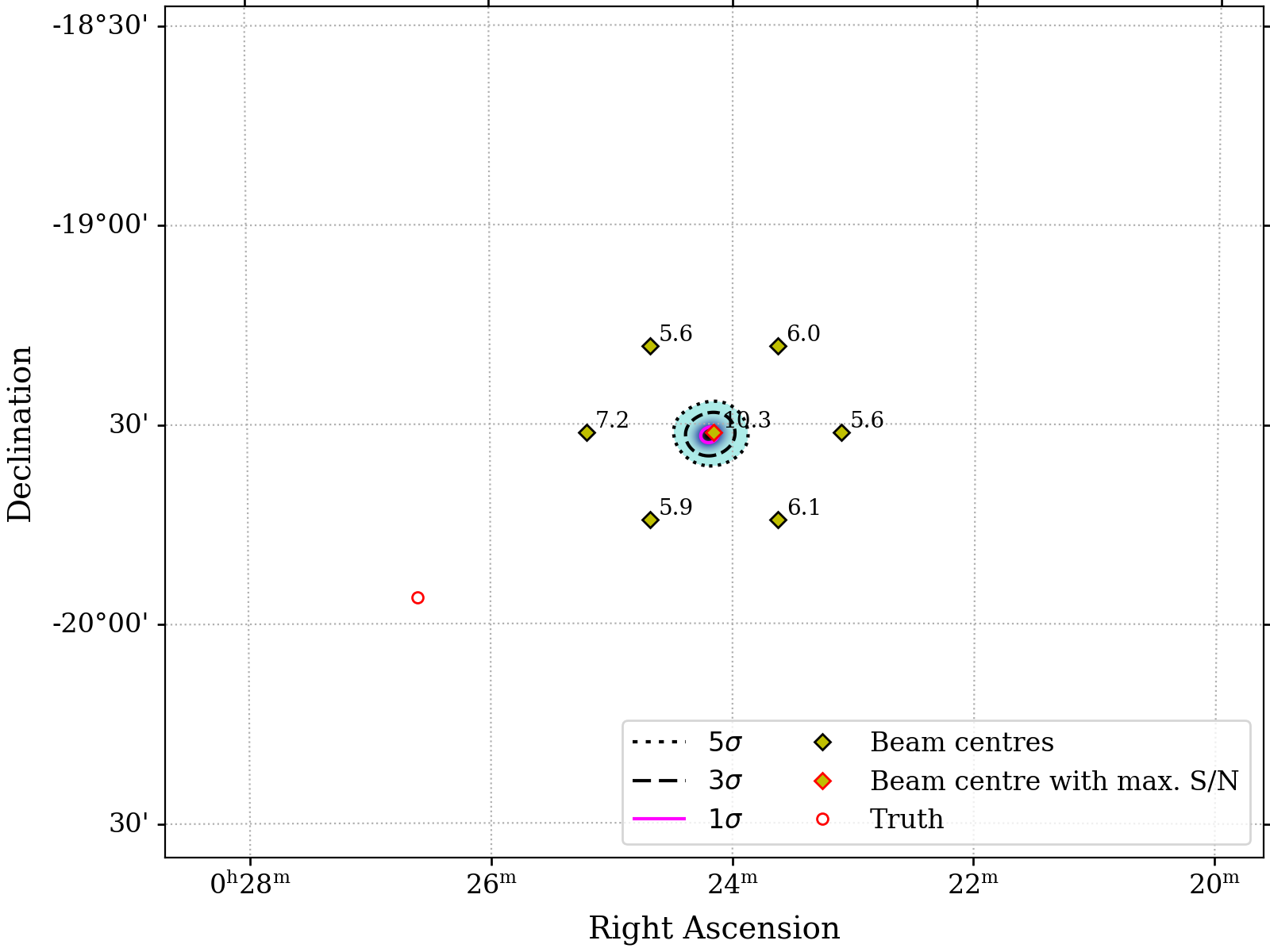}
    \caption{The localisation probability map computed over the same sky area for the same inputs based on an initial side-lobe detection of PSR~J0026$-$1955, with three different regularisation schemes. 
    The figure legend in the lower sub-figure applies to the others.
    \textit{Top}: Without any re-weighting of the statistical map. 
    \textit{Middle}: Regularised with a wide Gaussian window centred at the TAB with the largest detection S/N. 
    \textit{Bottom}: Regularised with the TAB pattern corresponding to the TAB with the largest detection S/N.
    Both the Gaussian regularised and un-regularised maps suggest that follow-up is required near the (ultimately correct, based on interferometric imaging) pulsar position, which would ultimately lead to more significant detections and a more confident localisation.}
    \label{fig:app-regularisation-diff}
\end{figure}

Here we compare what the localisation would have been with only the original side-lobe detection information. 
The comparison is visualised in Fig.~\ref{fig:app-regularisation-diff}.
As noted in the main text, it is important to always inspect the un-regularised localisation map, especially when there is no other \textit{a priori} information known about the object sky position.
Classical centroiding algorithms would also suffer from the same issue as the TAB-weighted case. 

\end{document}